\documentclass[aps,prb,twocolumn,groupedaddress,showpacs]{revtex4}
\usepackage{amsfonts}
\usepackage{mathrsfs}
\usepackage{graphicx}

\begin{document}

\title{Coherent dynamics of singlet fission controlled by nonlocal electron-phonon coupling}

\author{Yao Yao$^{1,2}$\footnote{Electronic address:~\url{yaoyao@fudan.edu.cn}}}

\affiliation{
$^1$State Key Laboratory of Surface Physics and Department of Physics, Fudan University, Shanghai 200433, China\\
$^2$Collaborative Innovation Center of Advanced Microstructures, Fudan University, Shanghai 200433, China\\
}

\date{\today}

\begin{abstract}
Based on the Frenkel-charge transfer (CT) mixing model of singlet fission (SF), we incorporate both the local and nonlocal phonon baths in the Hamiltonian and adopt the algorithm of time-dependent density matrix renormalization group to simulate the fission process in tetracene. The endergonic SF is found to be facilitated by the robust quantum coherence, which concurrently gives rise to a notable quantum beating effect. Controlled by the nonlocal electron-phonon coupling, the density of triplet yield manifests a nonlinear relationship with the singlet density. As a result, the fission rate in tetracene is explicitly obtained, which is in agreement with the experimental measurement.
\end{abstract}

\pacs{88.40.jr, 71.35.Cc, 71.38.-k}

\maketitle


Providing rapid advances of power conversion efficiency (PCE), photocells based on organic materials exhibit promising application potential \cite{bredas,CR}. From the viewpoint of heat engine model proposed by Shockley and Queisser, however, the excitation energy above the bandgap of the light harvesting material is largely dissipated, lowering the PCE in a considerable manner \cite{SQ}. A strategy to overcome this Shockley-Queisser limit of PCE relies on a mechanism of producing more than one electron/hole by a single photon, namely converting one exciton into two or more \cite{SFreview,SFreview2}. This is plausible considering the spin exchange energy between electron and hole may make the energy of singlet exciton about twice larger than that of triplet. Together with a moderate intermolecular transfer rate of electron/hole, there are good chances for a singlet exciton to split into two triplets. This so-called singlet fission (SF) process, whose first experimental prototype appeared in 1960¡¯s \cite{SF0}, was implemented in a large family of organic material, such as tetracene, pentacene and hexacene \cite{SF1,SF2,SF3,SF4,SF5,SF6}. The updated advances achieve near-unity quantum yield of triplets, which is also the theoretical limit of SF \cite{SF3}, so that it can serve as an inspiring source for sizably improving the PCE of organic photocells \cite{SFT1,SFT2,SFT3,SFT4,SFT5,SFT6,SFT7,SFT8,SFT9,SFT10}.

Intuitively, there are two competing underlying mechanisms for SF: one-step and two-step \cite{SFreview2}. The former mechanism calls for the superexchange of electron spin to directly convert the singlet exciton (S1) to two triplet excitons (TT). Due to the weak spin interaction in organic materials \cite{Dediu}, this effect does not exhibit visible effect. In an alterative way, the latter one infers to a mediated mechanism, that is, the S1 is firstly transformed into an intermolecular charge transfer (CT) state followed by a second step from CT state to TT state. In order to validate this mechanism, a recent experiment of two-photon photoemission spectra in the tetracene figures out the presence of high-efficiency endothermic SF \cite{SFT4}. The coherent superposition between singlet and CT state is addressed to explicate how the relatively high-energy CT and TT state participate in the fission process. Another experiment of nonlinear density dependence of singlet and triplet excitons further indicated the important role of wavefunction delocalization of electrons in the crystalline tetracene films \cite{SFT10}. Given these remarkable findings, the mediated mechanism is currently thought to be dominant, and the quantum coherence of CT state emerges as an irreducible ingredient in the endergonic SF process. However, due to the large energy gap between S1 and CT state (about 1eV) and the thermal dissipation process in the CT state, the quantum coherence is missing in some semiclassical theoretical methods such as the kinetic algorithm and the master equations formalism \cite{Th1,Th2,Th3,Th4}, and therefore it is not possible to calculate the endergonic SF based on these theoretical methods \cite{SFT8}. A full-quantum dynamical simulation is consequently required, which turns out to be the main goal of the present work.

Theoretically, a benchmarking model taking both the Frenkel exciton and the CT state into account has been established, and the density functional theory (DFT) calculations have correspondingly carried out the electron orbital couplings based upon the configuration interaction formalism \cite{Th2,Th3}. Rather, contentions surrounding the appropriate parameters of the model stay active \cite{Th1,Th2,Th3,Th4}. The key parameters such as the energy of the electronic state and the electron-phonon couplings are yet to be determined solely by DFT \cite{Th2}. Obtaining the energy of an electronic state requires the corresponding wavefunction, but the wavefunction of electron in the real materials is greatly delocalized as indicated by experiment \cite{SFT10}, which can not be handled by DFT with local functionals. The electron-phonon coupling matters in a dynamical fashion by affecting the decoherence rate and thus the rate of SF \cite{SF5}. In this context, one would think of employing the full-quantum algorithm to fit the relevant experimental results and then determine the associated microscopic parameters.

Our starting point is based on a recent experiment on the multiphonon relaxation in SF \cite{SF5}, which sheds light upon the theoretical study on phonons and thereby the quantum coherence. The model Hamiltonian is written as
\begin{equation}
H=H_{\rm el}+H_{\rm ph}+H_{\rm el-ph}.\label{hami}
\end{equation}
Herein, the first term $H_{\rm el}$ represents the Frenkel-CT mixing model for the molecular dimer \cite{Th2,Th3}, that is,
\begin{equation}
H_{\rm el}=\sum_i|i\rangle E_i\langle i|+\sum_{i\neq j}|j\rangle V_{ij}\langle i|,
\end{equation}
where $|i\rangle$ takes five possible states including two singlet states $|1\rangle\equiv|S_1,S_0\rangle$ and $|2\rangle\equiv|S_0,S_1\rangle$, two CT states $|3\rangle\equiv|C,A\rangle$ and $|4\rangle\equiv|A,C\rangle$ (C for cation and A for anion), and one TT state $|5\rangle\equiv|T_1,T_1\rangle$; $E_i$ is the respective energy of the state, and $V_{ij}$ is the transition energy from state $i$ to state $j$. As stated, the values of $V_{ij}$ have been obtained by DFT calculations \cite{Th2,Th3}, but $E_i$'s are yet to be determined.

More important are the terms involving phonons. The second and third term of Hamiltonian (\ref{hami}) read ($\hbar=1$),
\begin{eqnarray}
H_{\rm ph}&=&\omega_{\rm L}\hat{a}^{\dag}\hat{a}+\omega_{\rm NL}\hat{b}^{\dag}\hat{b},\\
H_{\rm ex-ph}&=&\gamma_{\rm L}|5\rangle\langle 5|(\hat{a}^{\dag}+\hat{a})\nonumber\\&+&\gamma_{\rm NL}(|3\rangle\langle 3|+|4\rangle\langle 4|)(\hat{b}^{\dag}+\hat{b}),\label{hamiexph}
\end{eqnarray}
where $\hat{a}^{\dag} (\hat{a})$ and $\hat{b}^{\dag} (\hat{b})$ are the creation (annihilation) operator of the local and nonlocal phonons with the frequency being $\omega_{\rm L}$ and $\omega_{\rm NL}$, and $\gamma_{\rm L}$ and $\gamma_{\rm NL}$ are the coupling strength, respectively. Here, we have two kinds of phonons, local and nonlocal phonons, which have been widely studied in organic crystalline materials \cite{ep0,ep1,ep2,ep3}. The local phonons couple with the tightly bounded excitons in a single molecule. In the current work, we consider the local phonons coupled with the TT state, which is the main concern in the SF process. The nonlocal phonons interact with the CT state, since by definition the CT state is an intermolecular electron-hole pair and the intermolecular vibrational modes take over the relevant coupling \cite{ep0}. As stated, the CT state plays an essential role in the SF process, implying the nonlocal phonons would be significant. For simplicity, we assume that both the local and nonlocal phonon baths follow the similar spectral density function, which is usually cut off at the frequency $\omega_c$, i.e., $J_{\rm L(NL)}(\omega)=2\pi\alpha_{\rm L(NL)}\omega^{1-s}_c\omega^{s} e^{-\omega/\omega_c}$ with $\alpha_{\rm L(NL)}$ being the dimensionless coupling strength for the local (nonlocal) baths and $s$ being the exponent. We mainly consider the sub-Ohmic bath $s=0.5$ corresponding to the relatively low-frequency vibrational modes, which are the most featured ones in organic molecules \cite{ep0,ep1}. The sub-Ohmic bath holds the non-Markovian feature to induce the long-lived quantum coherence \cite{mine2}. It is noticed that, the behavior of the SF process could be quantitatively different if other forms of spectral density were found to be appropriate.

Technically, dealing with two phonon baths in the meantime turns out to be significantly difficult in the framework of the full-quantum simulation \cite{mine1}. Recently, our effort based on the time-dependent density matrix renormalization group algorithm \cite{tDMRG} within the orthogonal polynomials representation \cite{Chin2,Chin1,Guo1,mine2,mine3} has come up with an \textit{ad hoc} optimizing method, so-called symmetrically optimized phonon basis \cite{mine4}, which allows us to simultaneously study the two phonon baths in a unified framework. In the following, we employ this method to calculate the dynamics of Hamiltonian (\ref{hami}) with the initial state being a singlet excitation $|1\rangle$. The parameters in tetracene are taken as \cite{Th3}: $V_{13}=V_{31}=-0.051$eV, $V_{14}=V_{41}=-0.074$eV, $V_{23}=V_{32}=-0.118$eV, $V_{24}=V_{42}=-0.111$eV, $V_{35}=V_{53}=-0.081$eV,$V_{45}=V_{54}=0.056$eV, and other $V_{ij}$'s are zero. $E_{\rm S1}(\equiv E_1=E_2)$ is set to be zero, and other parameters are adjustable in the computations.

\begin{figure}
\includegraphics[angle=0,scale=1.5]{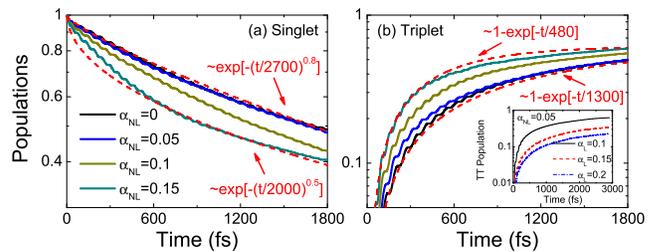}
\caption{Population evolution of (a) S1 and (b) TT state with four $\alpha_{\rm NL}$'s. The other parameters are: $E_{\rm CT}=0.6$eV, $E_{\rm TT}=0.1$eV, and $\alpha_{\rm L}=0.1$. The red dashed lines show the respective fitting curves (see in the text). The inset of (b) shows the population evolution of TT state for $\alpha_{\rm NL}=0.05$ and three $\alpha_{\rm L}$'s.}\label{fig1}
\end{figure}

We first show the population of S1 state ($|1\rangle$ plus $|2\rangle$) and TT state ($|5\rangle$) in Fig.~\ref{fig1}, with $E_{\rm CT}(\equiv E_3=E_4)=0.6$eV, $E_{\rm TT}(\equiv E_5)=0.1$eV, and $\alpha_{\rm L}=0.1$. In the present work, both the local and nonlocal dimensionless couplings $\alpha_{\rm L(NL)}(\sim \gamma_{\rm L(NL)}/2\pi)$ are taken to be around 0.1, since the effective value of Peierls coupling in pentacene is about 0.55 and we estimate it to be similar in tetracene \cite{ep0}. It is found that, the singlet population decays from one and the triplet population grows up from zero, implying the singlet is efficiently converted to the triplet, namely SF takes place. At the very initial stage, i.e. the time $t$ is smaller than 1000fs, a quantum beating behavior originated from the quantum coherence emerges for both the singlet and triplet population \cite{beat}. After that, the populations gradually saturate during a long-term evolution. It is worth noting that the quantum beating effect produced by our non-perturbative full-quantum simulation can not be obtained in the framework of incoherent hopping model with Redfield or F\"{o}rster theory \cite{ThNJP}. The quantum-mechanically treated phonons give rise to the essential and robust quantum coherence, which is crucial for the endergonic SF as discussed below.

\begin{figure}
\includegraphics[angle=0,scale=0.5]{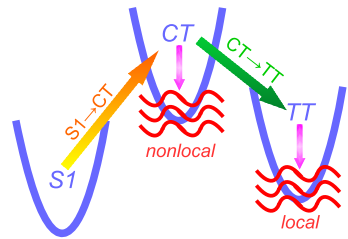}
\caption{Schematic for the SF process and the role of local and nonlocal phonons. The CT and TT state are coupled with the nonlocal and local phonon baths, respectively. The baths serve as the sink for the both processes from S1 to CT and from CT to TT.}\label{fig2}
\end{figure}

Significantly, we find the conversion from singlet to triplet is accelerated by the nonlocal phonons. We fit the curves of $\alpha_{\rm NL}=0$ and $0.15$ using the exponential function $\exp[-(t/t_0)^{\nu}]$ with $t_0$ being the characteristic fission time for the SF, and $\nu$ the exponent. For the singlet population, two functions $\exp[-(t/2700)^{0.8}]$ and $\exp[-(t/2000)^{0.5}]$ are assigned for the curves of $\alpha_{\rm NL}=0$ and $0.15$, respectively. It is clear that $t_0$ for the large nonlocal coupling case is shorter than that for the small coupling case, indicating the large nonlocal coupling gives rise to a large fission rate. Furthermore, the exponents for the two curves are quite different. When the nonlocal phonon is absent, the decay of singlet population approximately follows an exponential behavior, while when a large nonlocal coupling participates in, the exponent becomes much smaller than one. In order to clarify the physical meaning of these fitting parameters, we write down a diffusion equation for the density of singlet $\rho_{\rm S1}(t)$ \cite{SFT10},
\begin{eqnarray}
\frac{d\rho_{\rm S1}(t)}{dt}=-k_{\rm SF}\rho_{\rm S1}(t),
\end{eqnarray}
with $k_{\rm SF}(\sim t^{\nu-1})$ being the fission rate. Obviously, a unity $\nu$ gives rise to a normal diffusion behavior, while a smaller-than-unity $\nu$ refers to a sub-diffusion case. This anomalous diffusion effect, which would result in a nonlinear relationship between singlet and triplet, has been observed in the corresponding experiment \cite{SFT10}. It arises from the many-body interaction between electrons and the nonlocal phonons accompanied with the CT state. To be more elaborative, we draw a schematic in Fig.~\ref{fig2} to show the role of both local and nonlocal phonons. Right after the S1 is transferred to the CT state, the CT state could simultaneously dissipate the energy to the nonlocal phonon bath and recombine to the TT state. The former process could strengthen the latter because the nonlocal phonons acts as a sink of the CT state energy, which constantly breaks the detailed balance between the S1 and CT state population and drives the S1-to-CT state conversion, as shown in Fig.~\ref{fig1}(a). As the nonlocal phonons mainly refers to the intermolecular vibrational modes, an experimental corroboration for our present viewpoint is the fast charge transfer process in polar solvents \cite{ACR}. The weak covalent coupling helps form a stable dipolar intermediate and thus enhance the intermolecular couplings, inducing the large fission rate.

Fig.~\ref{fig1}(b) displays the production of triplet excitons. Different from that of singlet, the curves of triplet are well fitted by the function of $[1-\exp(-t/t_0)]$, with $\nu$ being one. This implies the nonlinear dependence of the triplet on the singlet population in a robust manner. More interestingly, $t_0$, the characteristic fission time, greatly differs from each other for different $\alpha_{\rm NL}$'s. For $\alpha_{\rm NL}=0$, $t_0\simeq1300$fs, while for $\alpha_{\rm NL}=0.15$, $t_0\simeq480$fs, about three times smaller than that in the former case. Subsequently, the larger the nonlocal coupling, the shorter the fission time. For comparison, the influence of local phonons is shown in the inset of Fig.~\ref{fig1}(b), where three $\alpha_{\rm L}$'s are calculated for fixed $\alpha_{\rm NL}=0.05$. It is very clear that all the three curves share the same fission time; they merely differ from each other on the saturated value of TT population. Compared with the corresponding experiment \cite{SF5}, it is the multiphonon relaxation that reduces the TT yield mostly. To summarize, the nonlocal phonon serves as a controller for the coherent SF process, which not only controls the fission rate but also influences the yield of triplet excitons.

\begin{figure}
\includegraphics[angle=0,scale=1.5]{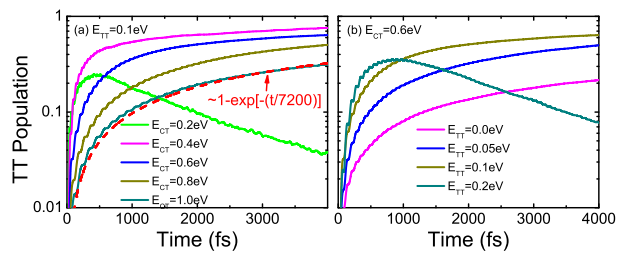}
\caption{Population evolution of TT state for various sets of energy of CT and TT state with (a) $E_{\rm TT}=0.1$eV and (b) $E_{\rm CT}=0.6$eV. The other parameters are: $\alpha_{\rm L}=0.1$, $\alpha_{\rm NL}=0.1$. The red dashed line in (a) shows the respective fitting curve.}\label{fig3}
\end{figure}

In the following, we discuss the endergonic SF process \cite{SFT4,SFT10}. The energy of both CT and TT state is changed (always above the energy of S1 state, 0eV) as shown in Fig.~\ref{fig3}. We first take $E_{\rm CT}=1.0$eV and $E_{\rm TT}=0.1$eV, the estimated values obtained by the DFT computations \cite{Th3}. The curve is fitted by the exponential function with the fission time $t_0$ being $7200$fs, in good agreement with the experimental measurement in tetracene \cite{note}. Moreover, when the energy of CT state is reduced, the fission rate is at the initial stage of the evolution enhanced, and finally the yield of triplet is increased by about twice. The energy barrier between S1 and CT state comes on duty in this effect. In addition, when $E_{\rm CT}$ is close to $E_{\rm TT}$ (e.g. $E_{\rm CT}=0.2$eV), the TT population exhibits a fast decay in the time evolution as it goes back to the CT state. Experimentally, the change of the CT state energy could be implemented by adjusting the excitation power \cite{SFT10}. Under higher density of excitons, more high-energy CT state could be occupied. In other words, the larger the power, the higher the energy of the highest occupied CT state \cite{entropy}. It has been observed that under larger power of excitation, the fission of singlet becomes weaker, so the yield of triplets becomes smaller \cite{SFT10}. Obviously, our calculated results match the experiment very well, and the fitting parameters employed here turn out to be approximately the associated parameters of the real material.

\begin{figure}
\includegraphics[angle=0,scale=0.6]{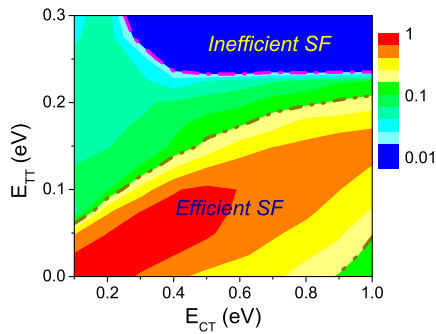}
\caption{Triplet population at $4000$fs versus the energy of both CT and TT state for $\alpha_{\rm L}=0.1$, $\alpha_{\rm NL}=0.1$. The dash-dot-dot and dash-dot boundary lines indicate the parameter regimes for efficient ($\rho_{\rm TT}>0.1$) and inefficient SF ($\rho_{\rm TT}<0.01$), respectively.}\label{fig4}
\end{figure}

The curves for various $E_{\rm TT}$ are shown in Fig.~\ref{fig3}(b). The increase of $E_{\rm TT}$ gives rise to the increase of triplet yield for $E_{\rm TT}\leq0.1$eV, otherwise the TT population decays. The SF is quite efficient as the saturated population of triplet for $E_{\rm TT}=0.1$eV is larger than 0.5. In Fig.~\ref{fig4} we further show the dependence of the triplet yield and the energetics of electronic states, that is, the population of TT state $\rho_{\rm TT}$ at $4000$fs versus the energy of both CT and TT state. To be clearer, we define two parameter regimes of \textit{efficient} and \textit{inefficient} SF to be those for $\rho_{\rm TT}>0.1$ and $\rho_{\rm TT}<0.01$, respectively. The location of the boundaries are qualitatively different from that calculated by the perturbative master equations approaches \cite{Th2,Th3}. Although the most efficient fission occurs in the regime of small TT energy, but remarkably, the TT energy for efficient SF could be larger than that of S1 state. Based on these results, we thus theoretically prove the hypothesis that, benefitting from the participation of quantum coherence, the superposition of S1 and CT state makes endergonic SF take place in an efficient fashion \cite{SFT4}.

In summary, the local and nonlocal electron-phonon couplings are incorporated into the Frenkel-CT mixing model, and a full-quantum algorithm is adopted to calculate the coherent dynamics of SF process. The quantum beating is obtained at the initial stage of the evolution, manifesting the proper consideration of quantum coherence. By enhancing the nonlocal coupling, both the SF rate and the triplet yield are increased. The nonlinear relationship between the populations of singlets and triplets is rebuilt, and the nonlocal coupling turns out to be the controller for these effects. The energetics of the electronic state is varied to fit the experimental measurement. The SF time of 7200fs in tetracene as well as the endergonic SF phenomenon are obtained by our simulations. In consequence, the present theoretical model has successfully mimicked the coherent SF process in tetracene, and we expect that it could provide some insights for its application on solar energy harvesting.

\begin{acknowledgments}
The author gratefully acknowledges support from the National Natural Science Foundation of China (Grant Nos.~91333202, 11574052, and 11134002) and the National Basic Research Program of China (Grant No.~2012CB921401). We are thankful to C. Zhang and R. Wang for useful discussions and W. Yang for proofreading the manuscript.
\end{acknowledgments}


\begin{thebibliography}{99}
\bibitem{note} The characteristic fission time is sensitive to the choice of experimental technique \cite{SFT7}. The photoemission spectroscopy gives the value 7.2ps \cite{SFT4}, while the transient photoluminescenece gives 110ps \cite{SFT7}. Theoretically, due to the influence of structural and thermal environment, the electron-phonon coupling is variable. By adjusting the coupling strength in a reasonable extent, it is possible to explicitly fit the measured values.
\end{thebibliography}
\end{document}